# The magnetocaloric and barocaloric effects in $Gd_5Si_2Ge_2$ compound

J. D. ZOU*

*School of Materials Science and Engineering, Beihang University, Beijing 100191, China*

**Abstract**

The caloric effect in $Gd_5Si_2Ge_2$ compound is studied experimentally and theoretically. The first-order phase transition is sensitive to both magnetic field and pressure. It indicates that the influences of magnetic field and pressure on the phase transition are virtually equivalent. Moreover, the theoretical analyses reveal that the total entropy change is almost definite at a certain Curie temperature whether the applied external field is magnetic field or pressure. On the basis of theoretical analyses, the entropy change simultaneously induced by magnetic field and hydrostatic pressure can be obtained experimentally. As for $Gd_5Si_2Ge_2$ compound, the entropy change curve is broadened dramatically, and the refrigerant capacity is improved from 284.7 to 447.0 J/kg. Additionally, the required magnetic field in magnetic refrigeration technique may be reduced by the application of hydrostatic pressure.

*E-mail: zoujd@buaa.edu.cn

Current address: The Ames Laboratory, US Department of Energy, Iowa State University, Ames, Iowa, 50011-3020, USA

E-mail: zoujd@ameslab.gov

Researches in magnetocaloric effect (MCE) have focused on the behaviors of materials in the vicinity of first-order phase transition. The discovery of giant MCE in $Gd_5Si_2Ge_2$ [1, 2] extensively promotes the research on magnetocaloric materials and magnetic refrigeration technique. Since then, many large magnetocaloric materials [3-13] and magnetic refrigeration prototype instruments [14-16] have been reported. MCEs originate from the change in the microscopic magnetic order induced by applied magnetic field, and are marked by the isothermal entropy change or adiabatic temperature change. The performance of a magnetic refrigerant relies on the size of MCE and the strength of magnetic field. It should be noted that the rather high magnetic field usually required by the magnetic refrigeration may impede the practical application to be realized simply and reliably. As for materials with large MCE, they have one common peculiarity that their first-order phase transitions are sensitive to several external parameters, such as magnetic field, temperature, and pressure etc. This feature may be of great benefit to improve the performance of magnetocaloric materials and develop new magnetic refrigeration instruments. Following discussion will reveal the potential advantages of magnetocaloric materials theoretically and experimentally, and develop a method to evaluate the caloric effect during a first-order phase transition under both magnetic field and hydrostatic pressure.

Note that all giant MCEs are observed in the vicinity of first-order phase transitions where the phase transition temperatures are very sensitive to the magnetic field or pressure. Indeed, these phenomena reflect the internal intrinsic coupling effect between the magnetic and structural degrees of freedom, termed magnetostructural (or magnetoelastic) coupling effect. This unique feature provides a scarce possibility by which the phase transition can be manipulated flexibly. As results, not only the applicabilities of magnetocaloric materials may be enhanced, but also the MCE and magnetic refrigerant capacity (RC) may be improved. Oliveira *et al* have theoretically studied the magnetocaloric and barocaloric effect (BCE) of $ErCo_2$ [17] and $La(Fe_{0.89}Si_{0.11})_{13}$ [18] under pressure and magnetic field, and revealed that the large caloric effect may be obtained in a wide range of temperature. The mechanism of BCE is similar with that of MCE described above, but the pressure instead of magnetic field. Mañosa *et al* reported the large BCE and MCE in Ni-Mn-In alloy [19] which may originate from discontinuities of volume and entropy. These results indicate that the magnetocaloric materials may be of great talent which does not be fully appreciated. Therefore, more effort should be devoted to investigate the mechanism of caloric effect in the vicinity of first-order phase transition. $Gd_5Si_2Ge_2$ is employed to investigate the behavior of caloric effect under magnetic field and hydrostatic pressure due to its typical first-order phase transition and giant MCE.

With increasing the concentration of Ge, the $Gd_5Si_{4-x}Ge_x$ compounds can form $Gd_5Si_4$-type structure, $Gd_5Si_2Ge_2$-type structure, and $Sm_5Ge_4$-type structure, and the phase transition temperatures are gradually lowered.[7, 20, 21] The phase transition of $Gd_5Si_{4-x}Ge_x$ compounds can be controlled by magnetic field, temperature, as well as pressure. [1, 2, 22, 23] Experimental studies show that the giant MCEs observed in $Gd_5Si_{4-x}Ge_x$ compounds are associated with a field-induced first-order structural transition from paramagnetic (PM) monoclinic to ferromagnetic (FM) orthorhombic structure. [24] Here we engage in thoroughly experimental studies of crystallographic, magnetic and caloric properties of $Gd_5Si_2Ge_2$ compound, allowing exploration of the mechanism of caloric effect. Moreover, general analysis will be carried out, and a possible theoretical model is developed and used to describe the behaviors of entropy change under both magnetic field and pressure.

$Gd_5Si_2Ge_2$ sample were prepared by arc melting the pure metals under an argon atmosphere in a water-cooled copper crucible. The purity of raw materials was 99.9% for gadolinium, was 99.99% for silicon and germanium. The sample was sealed in quartz tubes at high vacuum. The sample was annealed at 1393K for 14 days, and then quenched in liquid nitrogen. The temperature dependence of polycrystalline XRD data was collected by using Bruker D8 diffractometer. The dc magnetization and heat capacity measurements were performed using a physical properties measurement system (PPMS-9T, Quantum Design). The magnetic properties under hydrostatic pressure were measured by using magnetic properties measurement system (MPMS, Quantum Design) with a MPMS high pressure capsule cell (Honest Machinery Designer's Office, Japan). The pressure value was determined by pressure dependence of critical temperature of the superconducting state of Pb sensor placed inside the cell. All measurements were conducted after zero-field-cooling in fields up to 5 T.

The heat-treated $Gd_5Si_2Ge_2$ sample is monoclinic structure at room temperature [25]. Temperature dependence of x-ray diffraction (XRD) confirms the structural transformation from monoclinic to orthorhombic structure on cooling. [25] Previous reports have shown that this structural transformation accompanies with a magnetic transition from PM (monoclinic structure) to FM state (orthorhombic structure). [24] The Curie temperature of this sample is 268 K at 50 mT, [25] which is very near the previous report. [22, 26] This kind of transformation is very sensitive to the magnetic field and pressure, and referred to as magnetostructural phase transition. With increasing the magnetic field, the phase transition point linearly moves to higher temperature with an approximate slope of 5.7 K/T. The phase transition point is 272 K at 1 T, and reaches 295 K at 5 T. [25] The hydrostatic pressure also can linearly drive the phase transition point to higher temperature at an approximate slope of 4.8 K/kbar.[22]

The magnetostructural phase transition is a sign of the coupling effect existing between the magnetic and structural degrees of freedom. In this case, the external parameters, such as magnetic field or pressure, may actually play a role of trigger which can determine the phase transition occurring or not. Therefore the influence of magnetic field and pressure on the phase transition is virtually equivalent. Since the giant MCEs induced by magnetic field have been observed in $Gd_5Si_{4-x}Ge_x$ compounds, the giant BCE induced by pressure is naturally expected. However, the BCE cannot be obtained with ease due to the difficulty in manipulating pressure and detecting volume change during a measurement. Here we try to describe a simple method by which the caloric effect can be evaluated under magnetic field and pressure.

Considering the FM-PM phase transition, there are two cases as schematically demonstrated in fig. 1. First, the phase transition point is driven up to a higher temperature by the magnetic field, but is depressed to a lower temperature by the pressure, such as $LaFe_{13-x}Si_x$, [27-31], and MnAs. [3, 6] Second, the phase transition point is promoted to a higher temperature by both magnetic field and pressure, such as $Gd_5Si_{4-x}Ge_x$ [22, 23] and $La_{0.69}Ca_{0.31}MnO_3$ [32]. Figure 1(a) shows the possible spontaneous magnetization and Curie temperature $T_C$ varying with temperature under the magnetic field and hydrostatic pressure. The influence of magnetic field on $T_C$ is opposite to that of pressure. It is well known that both entropy and spontaneous magnetization can scale the order of the magnetic system. The corresponding entropy changes under magnetic field will, without doubt, counteract with those under pressure, shown in fig. 1(b). While the entropy changes under magnetic field will superpose with those under pressure in the second case. Figures 1(c) and 1(d) schematically show the possible spontaneous magnetization and entropy changes varying with temperature. The higher the Curie temperature shifts, the wider the entropy change curve is. Certainly, the actual shape of entropy change curve will depend on the actual size of magnetic field and pressure. Based on this feature, if the pressure is applied together with magnetic field, the magnetic field required in practical application may be reduced dramatically, while the magnetic RC is retained, and vice versa. This method will enhance the applicabilities and maneuverabilities of magnetocaloric materials, especially for $Gd_5Si_{4-x}Ge_x$ compounds. Usually, the entropy change curve as shown in figs. 1(b) and 1(d) cannot be obtained directly because the pressure and volume change cannot be obtained easily. Therefore, it is necessary to develop a simple yet effective theoretical model and experimental method which can be used to describe the behavior of caloric effect during a first-order phase transition under both magnetic field and pressure.

We have previously discussed the mechanism of the entropy change under magnetic field. [10-12] Here we will expand the discussion concerning about the behaviors of entropy change impacted by both magnetic field and hydrostatic pressure. According to Maxwell relations, [10-12] if the magnetic system is independent, the magnetic entropy change under an applied magnetic field can be expressed as

$$\Delta S = S(H_1, T) - S(H_0, T) = \int_{H_0}^{H_1} \left(\frac{\partial M}{\partial T}\right)_H dH \tag{1}$$

Equation (1) is widely used to calculate the entropy change under magnetic field, and can be integrated numerically. Similarly, if the structural system is independent, the baric entropy change under applied pressure can be expressed as

$$\Delta S = S(P_1, T) - S(P_0, T) = \int_{P_0}^{P_1} \left(\frac{\partial V}{\partial T}\right)_P dP \tag{2}$$

In fact, the eq. (2) is difficult to be calculated which has been mentioned above. It is well known that the first-order magnetic phase transitions are often companied by volume deformation even structural transformation. It indicates that there are strong coupling effects (such as magnetostructural or magnetoelastic coupling), herein both the magnetic and structural systems are not independent. Therefore the results calculated from eqs. (1) or (2) are virtually the total entropy change of the system, not sole magnetic entropy change neither baric entropy change. It means that the concept of MCE or BCE in the vicinity of first-order phase transition is far beyond the scope of traditional concept. In this case, the amount of total entropy change is almost definite at a certain Curie temperature whether the external field is magnetic field or pressure, written as

$$\Delta S(T_C, H) \cong \Delta S(T_C, P) \tag{3}$$

This relation is criterion what the following experimental method should be met.

Based on the above discussion, the entropy change curve as shown in fig. 1 (d) can be obtained experimentally. The magnetization isotherms of $Gd_5Si_2Ge_2$ between 256 and 300 K are shown in fig. 2(a). The magnetization isotherms at 4.22 kbar are shown in fig. 2(b). The entropy change $\Delta S$ as a function of temperature and magnetic field are calculated from the magnetization isotherms, and shown in fig. 3(a). The great "spike-like" peaks are observed. The mechanism of "spike-like" peaks appearing in the vicinity of first-order phase transition has been previously revealed. [12] The peak value of $\Delta S$ reaches 40.2 J/kgK at 0-5 T. Compared with the sharp peak, the smooth plateau of $\Delta S$ are almost half of the peak, and reach 19.0 J/kgK at 0-5 T. The entropy change determined by latent heat measurement is 18.9 J/kgK [25] which is nicely consistent with the plateau of $\Delta S$ 19.0 J/kgK. This result also dovetails to the previous report very well. [1] It indicates that the plateau of $\Delta S$ may be reflected the actual entropy change during the first-order phase transition. Compared with the results under magnetic field, the entropy change peaks move up to higher temperature under both magnetic field and hydrostatic pressure, but the utmost magnitude is decrease and reaches 21.2 J/kgK as shown in fig. 3(b). The smooth plateau of $\Delta S$ is 12.7 J/kgK at 0-5 T. The Curie temperature is 295 K at 5 T, and is 294 K at 4.22 kbar. Because of the shifts of Curie temperature are almost the same either at 5 T magnetic field or 4.22 kbar hydrostatic pressure, therefore the entropy change curve [such as shown in fig. 1(d)] can be obtained through superimposed effect. According to eq. (3), the entropy change at 0-5 T and 4.22 kbar are approximately equivalent to that at 0-4.22 kbar and 5 T. The entropy change at 0-5 T and 0-4.22 kbar, therefore, is obtained by superposing the curve at 0-5 T in fig. 3(a) with the curve at 0-5 T and 4.22 kbar in fig. 3(b), and shown in fig. 4. Comparing with results under magnetic field, the width of the entropy change curve is broadened dramatically when the hydrostatic pressure is applied at the same time. This is a convincing evidence to prove the large MCE and BCE in $Gd_5Si_2Ge_2$ compound. The magnetic RC is important to evaluate the performance of magnetocaloric materials. Here the magnetic RC is calculated in an optimum refrigeration cycle. [33, 34] The results show that the magnetic RC is remarkably improved from 284.7 J/kg to 447.0 J/kg. It is obvious that a rather prominent magnetic RC can be obtained with a lower magnetic field cost by applying pressure simultaneously which will be of great advantage for magnetic (or baric) refrigeration application.

The behavior of first-order magnetostructural phase transition impacted by magnetic field and hydrostatic pressure is analyzed theoretically and experimentally. A simple principle is proposed and employed to discuss the mechanism of MCE and BCE during a magneto-structural phase transition. The workable method is developed to deduce the entropy change under magnetic field and pressure, and applied to the discussion of $Gd_5Si_2Ge_2$ compound. The prominent caloric effects of $Gd_5Si_2Ge_2$ are confirmed under magnetic field and hydrostatic pressure. This method can not only reduce the strength of required field (such as magnetic field or pressure), but also dramatically broaden the width of the entropy change curve leading to greatly enhance the magnetic RC.

This work has been supported by the National Natural Science Foundation of China (Grant No. 50801015, and 50921003), and the Fundamental Research Funds for the Central Universities (Grant No. YWF-11-03-Q-003).

**Figure captions:**

**FIG. 1**. Schematic diagram of temperature dependence of spontaneous magnetization and entropy change. (a) The phase transition temperature is promoted to a higher temperature by magnetic field (red line), while is depressed to a lower temperature by hydrostatic pressure (blue line). (b) The corresponding entropy change under magnetic field and hydrostatic pressure during a phase transition as shown in (a). (c) The phase transition temperature can be promoted to a higher temperature by both magnetic field and hydrostatic pressure. (d) The corresponding entropy change under magnetic field and hydrostatic pressure during a phase transition as shown in (c).

**FIG. 2.** (a) Isothermal magnetizations of $Gd_5Si_2Ge_2$ compound versus applied field, where the magnetizations were measured at increasing magnetic field and ambient pressure. The ambient pressure is negligible during the magnetic measurement in PPMS. (b) Isothermal magnetizations versus applied field, where the magnetizations were measured at increasing magnetic field and 4.22 kbar hydrostatic pressure.

**FIG. 3.** Temperature dependences of total entropy changes. (a) The entropy change versus different magnetic fields. (b) The entropy change versus different magnetic fields at 4.22 kbar.

**FIG. 4.** Temperature dependences of total entropy changes at 0-5 T and 0-4.22 kbar. The magnetic RC is 284.7 J/kg at 0-5 T (as shown by the inner dash line), and is improved up to 447.0 J/kg at 0-5 T and 0-4.22 kbar (as shown by the outer dash line).

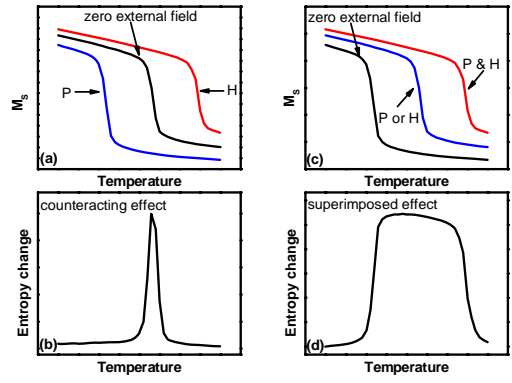

FIG. 1

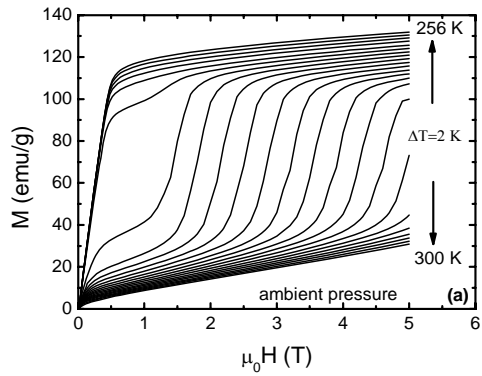

FIG. 2(a)

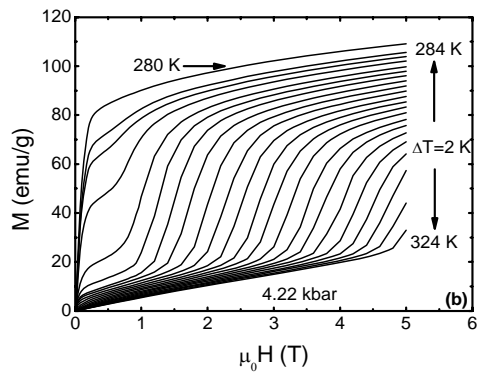

FIG. 2(b)

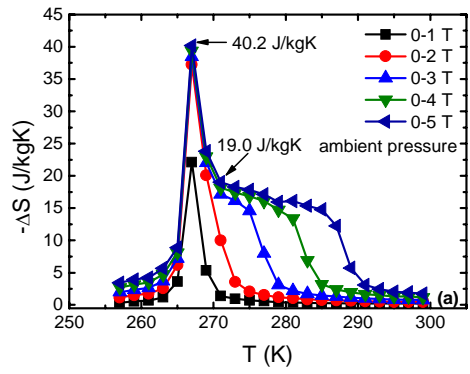

FIG. 3(a)

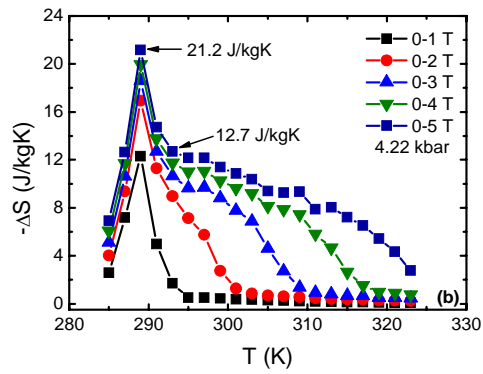

FIG. 3(b)

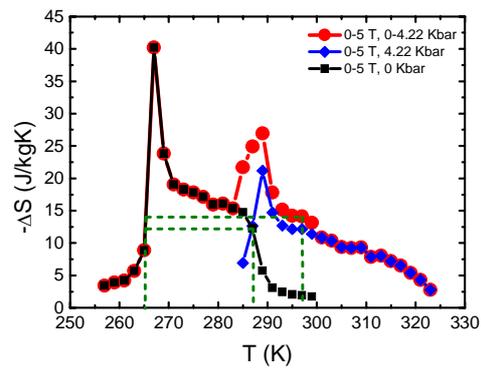

FIG. 4

Supporting materials for:

**The magnetocaloric and barocaloric effects in $Gd_5Si_2Ge_2$ compound**

J. D. Zou*

*School of Materials Science and Engineering, Beihang University, Beijing 100191, China*

*E-mail: zoujd@buaa.edu.cn

Current address: The Ames Laboratory, US Department of Energy, Iowa State University, Ames, Iowa, 50011-3020, USA

E-mail: zoujd@ameslab.gov

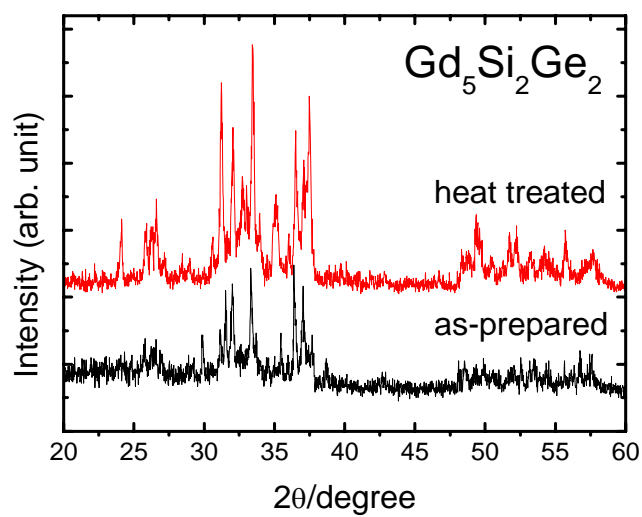

**FIG.S1.** XRD before and after heat treatment measured at room temperature.

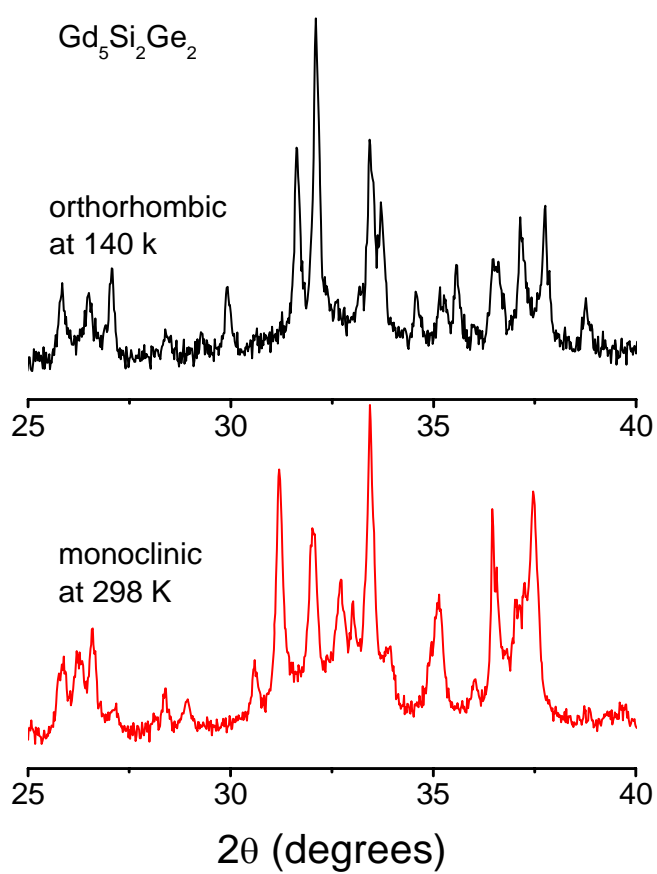

**FIG. S2.** XRD measured at 140 and 298 K, respectively. The sample changes from orthorhombic to monoclinic structure on heating.

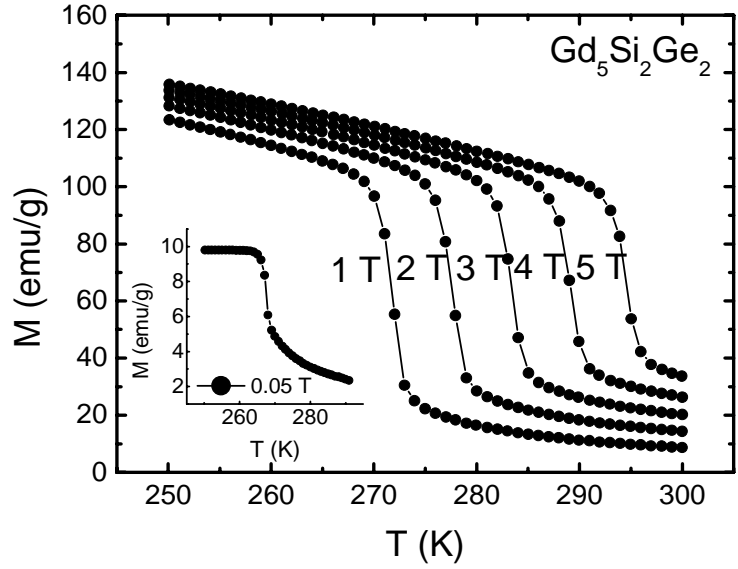

**FIG. S3.** Temperature dependence of magnetization under variable magnetic fields.

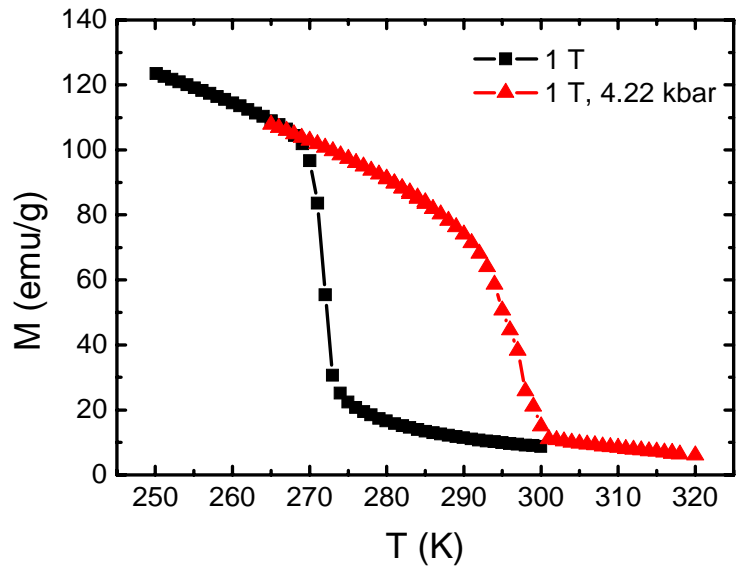

**FIG. S4.** Temperature dependence of magnetization. $T_C$ is 272 K at 1 T, and further moved up to 295 K when 4.22 kbar hydrostatic pressure is applied.

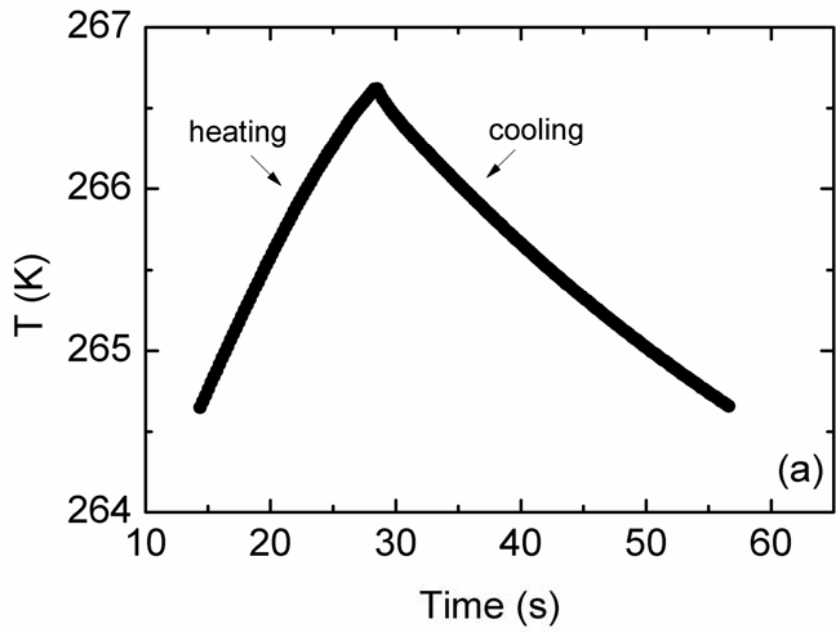

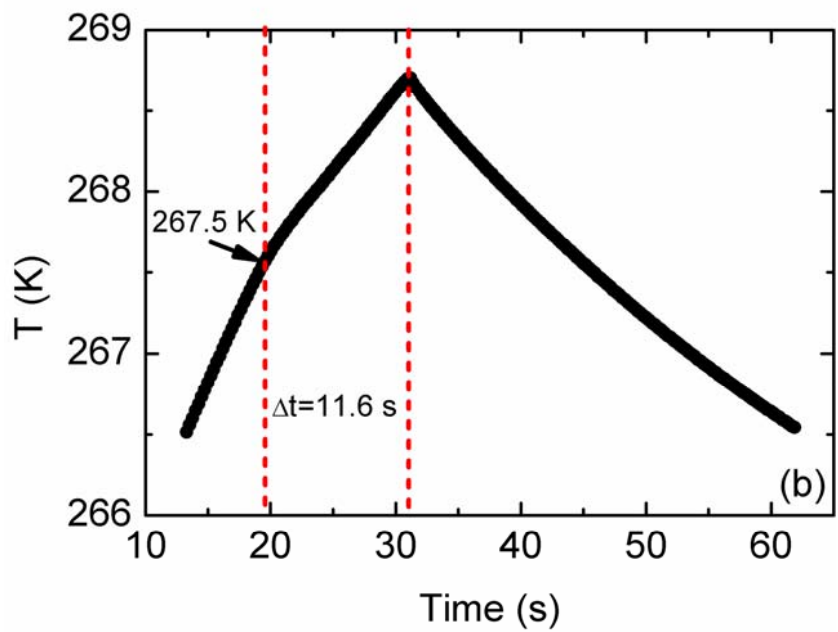

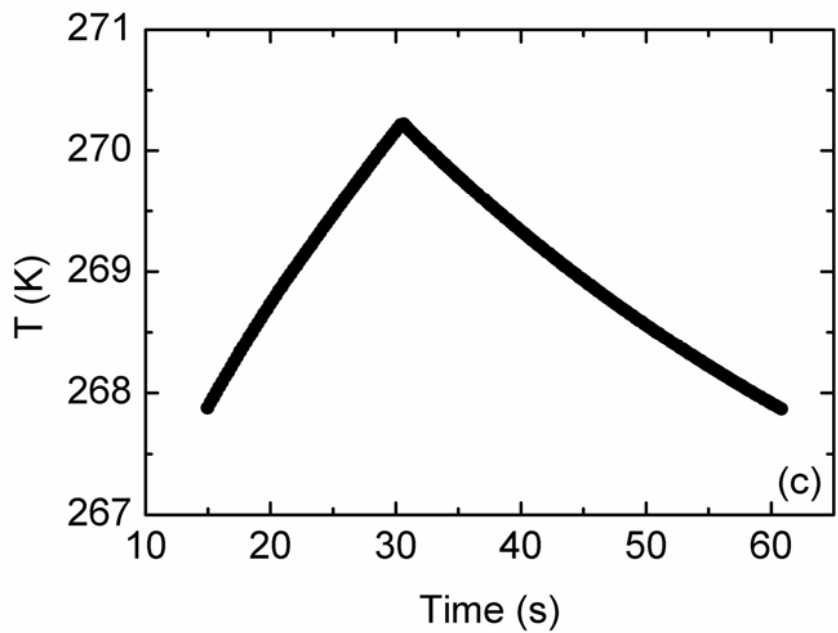

**FIG. S5.** It is well known that a first-order phase transition will accompany emission or absorption latent heat. According to thermodynamics theory, the entropy change is given by $dS=\delta Q/T$. Here $S$ is entropy, $Q$ is heat, and $T$ is temperature. Note that the $Q$ will equal to the latent heat at the critical temperature of a first-order phase transition. Then the entropy change can be expressed as $\Delta S=\Delta Q_L/T$, $Q_L$ is latent heat. Therefore, the entropy change during a first-order phase transition can be obtained by detecting the latent heat. The physical properties measurement system from Quantum Design can be used to measure the latent heat direct and credibly. Here we show the raw temperature response data near a first-order phase. (a) The heating and cooling curves before phase transition. (b) Whole first-order phase transition process. (c) The heating and cooling curves after phase transition. As shown in (b), the red dash lines mark the region of phase transition. The fixed heating power $P$ is 2.84 mW, and the duration of phase transition $\Delta t$ is 11.6 s, the latent heat $\Delta Q_L=P\times\Delta t=32.944$ mJ. Based on the sample mass of 6.5 mg and $T_C$ of 267.5 K, the entropy change during the first-order phase transition $\Delta S=\Delta Q_L/mT_C=18.9$ J/kgK which is nicely consistent with the plateau of $\Delta S$ 19.0 J/kgK determined by magnetic measurement. [As for this method, one may make reference to the manual of PPMS or Li *et al*, Phys. Rev. B 79, 054503 (2009)].

18